# FIELD DEPENDENCE OF THE INTERFACE ENERGY IN AF/AFM BILAYERS

J.R.L. de Almeida, J.R.Steiner, S.M.Rezende[*]

*Departamento de Física, Universidade Federal de Pernambuco, 50670-901, Recife-PE, Brazil*

**Abstract**

In the investigations of antiferromagnetic (AF)/ ferromagnetic (FM) bilayer samples, often distinct experimental techniques yield different values for the measured exchange anisotropy field ($H_E$). Using a simple microscopic model for representing the AF/FM interface, which incorporates the effect of interface roughness, we propose that the observed discrepancy may be accounted for by the dependence of the interface energy between the AF and FM layers with the value of the external applied field ($H$) as recently observed in anisotropic magnetoresistance measurements, lending support to our proposal.

It is well known that in the investigation of antiferromagnetic(AF)/ferromagnetic(FM) bilayer samples, often distinct experimental techniques yield different values for the measured exchange anisotropy field ($H_E$). This intriguing fact has been interpreted as arising from the distinct natures of the experimental techniques, some probe reversible properties of the system while others probe irreversible properties. Here we propose another reason for the observed discrepancy, namely the dependence of the interface energy between the AF and FM layers with the value of the external applied field ($H$), and the fact that each experimental technique employs a different field range. We use a simple microscopic model for representing the AF/FM interface, which incorporates the effect of interface roughness. This model has been used earlier to study the thermal-history-dependent properties observed in exchange-coupled AF/FM bilayers [1]. The results clearly demonstrate that the model captures the irreversibility and metastability properties of AF/FM bilayers, characteristic of spin-glasses and random-field systems, as observed experimentally [2]. The model also shows that the interface energy, which represents the exchange anisotropy field $H_E$, varies with the applied external field, as recently observed in anisotropic magnetoresistance measurements [3], lending support to our proposal. Consider two atomic monolayers with magnetic moments over congruent square lattices, one layer with ferromagnetically coupled moments and the other with two perfectly compensated antiferromagnetic sublattices. The moments from different layers are coupled by an interlayer exchange interaction, which can be FM or AF. The interface roughness is accounted for by randomly substituting a fraction of the atoms in the FM layer by atoms from the AFM layer. The system Hamiltonian is taken as

$$H = H_{AF} + H_{FM} + H_C$$

where $H_{AF}$, $H_{FM}$ and $H_C$ are, respectively, the interaction energies in the antiferromagnetic layer (AFML), in the ferromagnetic layer (FML) and the coupling between the FML and AFML atoms. As shown in [1] the interlayer exchange energy $H_c$ may be written as

$$H_C = -\sum_i \left[ J_c \eta_i S_i \sigma_i^{(1)} - J_1(1-\eta_i)\sigma_i^{(1)}\sigma_i^{(2)} \right]$$

where the sum is over all sites at the interface and $Jc$ represents the coupling between FML and AFML atoms, $\sigma_i^{(1)}$ means the spins on the AFML at site i, $S_i$ means the spins on the FML at site $i$, $\sigma_i^{(2)}$ denotes the moment of an AFM atom in FM layer and $\eta_i=1,0$ specify the presence (=1) or absence (=0) of a FML atom at site $i$ which, in the latter case, is assumed substituted by a AFML atom.

---

[*] Corresponding author e-mail: smr@df.ufpe.br

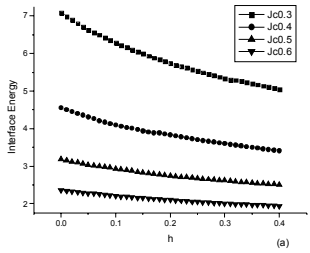

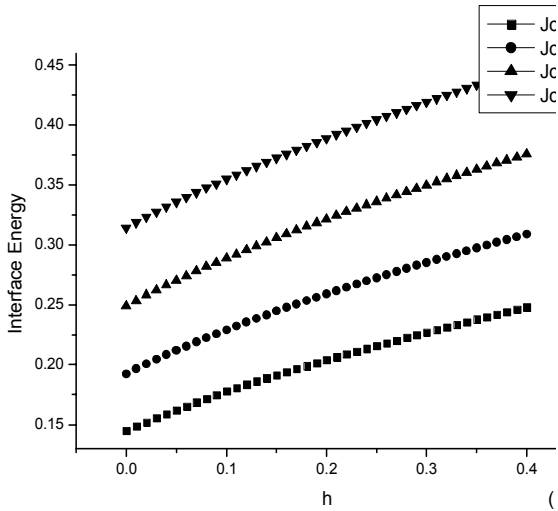

Fig. 1: Interface energy between the AF and FM layers as function of the applied field for S=1 systems.

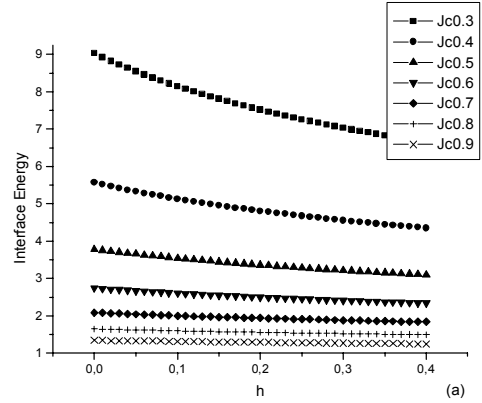

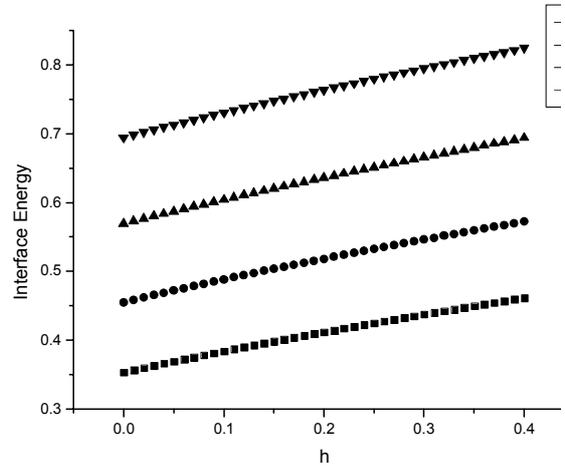

Fig. 2: Interface energy between the AF and FM layers as function of the applied field for S=2 systems.

All results presented here were obtained for two square lattice of size 2x100x100 with free boundary conditions, considering initially an AF/FM bilayer with AFM exchange $J_1$=-1.00 (energy scale) and roughness parameter $p$=0.30. In Fig. 1 and Fig. 2 we can see that the interface energy as function of the applied field is how observed experimentally in [3]. The new theoretical results are showed in Fig. 1 (b) and Fig. 2 (b), where we have the prediction of an inversion of behavior of the interlayer energy when the interlayer coupling is varied.